\documentstyle[sprocl]{article}

\def\Journal#1#2#3#4{{#1} {\bf #2}, #3 (#4)}

\def\PRD{{\em Phys. Rev.} D}

\def\be{\begin{equation}}
\def\ee{\end{equation}}
\def\bea{\begin{eqnarray}}
\def\eea{\end{eqnarray}}

\begin{document}

\title{RELATIVISTIC TRANSPORT EQUATIONS 
WITH GENERALIZED MASS SHELL CONSTRAINTS} 

\author{L. MORNAS}

\address{Departamento de F{\'\i}sica Te\'orica, Universidad de Valencia
\\ 46100 Burjassot (Valencia) Spain \\E-mail: mornas@chiral.ific.uv.es} 

\author{K. MORAWETZ}

\address{LPC/ISMRA, 6 Bd du Mar\'echal Juin, 14050 Caen, France
\\ E-mail: morawetz@caeau1.in2p3.fr}


\maketitle\abstracts{
We reexamine the derivation of relativistic transport equations for fermions 
when conserving the most general spinor structure of the interaction and
Green function. Such an extension of the formalism is needed when dealing
with {\it e.g.} spin-polarized nuclear matter or non-parity conserving
interactions. It is shown that some earlier derivations can lead to an 
incomplete description of the evolution of the system even in the case 
of parity-conserving, spin-saturated systems. The concepts of kinetic 
equation and mass shell condition have to be extended, in particular both 
of them acquire a non trivial spinor structure which describe a rich 
polarization dynamics.}

\vskip -0.8cm
\section{Introduction}
\vskip -0.1cm \noindent

We investigate how the spinor structure of the distribution function 
affects the transport properties of relativistic fermions. 
The generalized Wigner distribution function is the Green's function 
given by  
\vskip -0.4cm
\bea
F(x,p)= \int d^4 r\, e^{-i p.r}\, <{\overline {\Psi}}(x+{r\over 2}) 
\Psi(x-{r\over 2} )> .
\eea
\vskip -0.1cm \noindent
It can be developed on the basis of the Dirac gamma matrices as
\vskip -0.4cm
\bea
F(x,p)=f I + f_{\mu} \gamma^{\mu} + i f_{\mu\nu} \sigma^{\mu\nu}
+f_{5\mu} \gamma_5 \gamma^{\mu} +i f_5 \gamma_5.
\eea
\vskip -0.1cm \noindent
Its 16 components generally obey coupled transport equations (see {\it e.g.} 
\cite{FloHueKl96,MrHe94,HaMo92}). The usual Boltzmann equation for a
scalar distribution appears only if drastic assumptions are made.

The spin degrees of freedom in the context of nuclear matter, for example, 
might not have been given all the attention they deserve. Relying on 
the ``spin saturation'' argument, the derivation of transport theories 
for numerical simulations of heavy ion collisions generally assumes 
that the Wigner function is of the simpler form 
$F(x,p)=(\gamma^{\mu} . p_\mu + m) \varphi(x,p)$,  
which permits to recover the Boltzmann-Uehling-Uhlenbeck (BUU) 
kinetic equation. In this way, the only manifestation of the fermionic 
nature of the nucleons comes from the use of a Fermi-Dirac distribution 
function in the scalar $\varphi(x,p)$.

While the usual BUU approach is already satisfactory {\it e.g.} for the 
description of the gross features of heavy ion collisions, and has provided 
valuable insights in the interpretation of experimental data, we 
nevertheless believe that the spinor structure should be examined in
more detail. Our motivations are fourfold:
\newline \phantom{aaa} {\it (i)}
On the basis of a simplified relaxation time picture\cite{HaMo92}, 
a closed system of coupled equations for the scalar distribution $f$
{\it and} a spin distribution $f_{5\mu}$ was derived, and it was 
suggested that the nuclear plasma formed in a collision could relax 
more efficiently by exploring intermediate polarized states. On the 
other hand, the improvements brought lately in highly polarized beams 
and targets could be used to create a polarized excitation, and bring 
a new wealth of information on the dynamics of dense fermionic systems.
\newline\phantom{aaa} {\it (ii)}
It is interesting to remember at this stage the analogy with the
case of the spin-polarized helium fluids, as they provide an experimental 
check of the validity of the nonrelativistic limit of a transport 
theory with internal degrees of freedom \cite{helium}. There, 
effects such as increase of the viscosity, spin-heat coupling, 
spin waves in the non-degenerate regime can be observed
and their magnitude  can be compared to predictions.
Similar properties can be calculated in the nuclear system
\cite{HaMo92}.
\newline \phantom{aaa} {\it (iii)}
The interest in such studies has been revived in the context of
the study of the chiral transition, using the Nambu-Jona-Lasinio 
as a toy model for the QCD \cite{FloHueKl96,Flo98}. In the presence
of a pseudoscalar condensate, the assumption $F(x,p)=(\gamma^{\mu} . 
p_\mu + m) \varphi(x,p)$ cannot be made, instead a coupled transport 
equation for the system $f$, $f_{5\mu}$ is obtained.
\newline \phantom{aaa}  {\it (iv)}
Besides the transport equation, the relativistic formalism provides
also a constraint or mass shell equation. In the simple cases
where the spin saturation hypothesis is made, and also in the
chiral model considered by \cite{FloHueKl96,Flo98}, this equation 
takes the trivial form $p^2=m^2$.  In the general case however, 
the constraint equation also retains a spinor structure. One should
therefore be careful to use this constraint and not impose
from outside the usual $p^2=m^2$ condition for all components
of the Wigner function.
 
The derivation of the transport equations can proceed with the 
standard Green function formalism along the Keldysh 
contour. In a first step, we will assume that the particles are 
only interacting {\it via} the self-consistent mean-field 
$\Sigma_{MF}(X)=\Sigma$. The transport equations for the Wigner 
function reduce to 
\footnote{At this level of approximations, we have the properties:
$\gamma^0 F^\dagger \gamma_0 = F$ and $\gamma^0 \Sigma^\dagger \gamma_0 
= \Sigma$}:
\vskip -0.4cm
\bea
[ (i/2) \gamma.\partial + (\gamma.p -m) 
-\Sigma e^{-i \Delta \over 2} ] F(x,p) = & 0 \\
F(x,p) [ (i/ 2) \gamma.\partial - (\gamma.p -m) +
e^{i \Delta \over 2} \Sigma ] = & 0
\eea
\vskip -0.1cm
Here, $\Delta$ is the differential operator
$\Delta = \overleftarrow \partial_x . \overrightarrow \partial_p$.
This provides a complete system of differential equations describing 
the evolution of $F(x,p)$. In principle we could use equations 
(3,4) as they are.  We would like however to bring them in a
form which we could readily interpret physically as the Vlasov kinetic
equation plus a mass shell constraint.

The mass shell constraint is often considered as obvious, to the extent 
that some authors simply introduce it by hand at the end of the derivation 
of the kinetic equation. We would like to point out in this contribution 
that the mass shell equation is in general not a trivial one, but contains 
important dynamical and spinor information. In a first part, we show on 
the example of a simple model with scalar and vector mean fields, how 
different approaches could lead to different kinetic equations if the 
mass shell constraint is not taken consistently into account. In 
a second part, we compare the solution of the stationary problem for
pseudoscalar or pseudovector mean fields respectively, and illustrate 
on these examples that the mass shell constraint acquires a non trivial 
spinor structure.

\section{The $\sigma$-$\omega$ model of nuclear interactions and the
mass term issue}

We first consider the QHD-I model of the nuclear interaction\cite{SW86}
where we have scalar and vector mean fields only $\Sigma=(\Sigma_s 
+\Sigma_{v\mu} \gamma^\mu)$.

\subsection{Vlasov equation from the ``quadratic'' method}
\label{sec:qua}

A method to extract ``kinetic'' and ``constraint'' equations is 
that used by {\it e.g.} Elze et al \cite{EG87}:
The equations (3,4) are put in the form 
\vskip -0.4cm
\bea
& (\gamma.K-{\cal M})F=0 \qquad ; \qquad F(\gamma.K^\dagger-{\cal M}^\dagger)=0
\eea
\vskip -0.1cm \noindent
with ${\cal M}=M +i N$, $M=m-\Sigma_s$,
   $N={1 \over 2} \partial_\alpha(\Sigma_s) \nabla^\alpha$, 
$K^\mu=\Pi^\mu +{i \over 2} {\cal D}^\mu$, 
  $\Pi^\mu=p^\mu+\Sigma_v^\mu$, ${\cal D}_\mu=\partial_\mu 
  -\partial_\alpha{\Sigma_{v\mu}}\nabla^\alpha$, $\nabla^\alpha=\partial/
\partial  p_\alpha$.
By multiplying on the left hand side by $(\gamma.K+{\cal M})$ and on 
the right hand side by $(\gamma.K^\dagger+{\cal M}^\dagger)$,
and taking the sum and difference of the resulting equations, one obtains
in the first order of the gradient expansion 
\footnote{Terms such as ${\cal D}.{\cal D}/4 $ can be neglected.
The same argument can be given performing a semiclassical expansion
in powers of $\hbar$}, after some algebra
\vskip -0.4cm
\bea
& \left[ \Pi^\mu {\cal D}_\mu -M \partial_\alpha (\Sigma_s) \nabla^\alpha 
\right] F  +{1 \over 4} {\cal F}_{\mu\nu} \left[ \sigma^{\mu\nu}, F 
\right] +{1 \over 2} \partial_\mu(\Sigma_s) 
\left\{ \gamma^\mu, F \right\} = {\cal O}^2 &  \\
&\left[ \Pi^2-M^2  \right] F + {i \over 4} {\cal F}_{\mu\nu} 
\left\{ \sigma^{\mu\nu}, F \right\} +{i \over 2} \partial_\mu(\Sigma_s) 
\left[ \gamma^\mu, F \right] = {\cal O}^2 &
\label{7}
\eea
\vskip -0.1cm \noindent
where ${\cal F}^{\mu\lambda}=\partial^\mu(\Sigma_v^\lambda)
-\partial^\lambda(\Sigma_v^\mu)$.
In this way, familiar looking mass shell condition and 
kinetic equation are recovered.
This ``quadratic'' method however has been criticized \cite{Smol} on the 
ground that information could be lost in the process, in the same way as 
information is lost on the fermionic nature of the particle by going from 
the Dirac to the Klein Gordon equation.
The issue is important since these authors argue that an other  -- 
``non-quadratic'' -- method of derivation yields a different kinetic
equation where the mass term $M \partial_\alpha(\Sigma_s) \nabla^\alpha$ 
is absent. We will return to this issue in the next section. 
We first analyze more carefully the equations and show that the mass 
term is also obtained without squaring the Dirac equation.

\vskip -0.3cm \noindent
\subsection{A ``pedestrian'' method}
\label{sec:ped}

The components of system of equations (3,4) on the basis of 
Dirac gamma matrices yield in the first order gradient approximation 
\vskip -0.4cm
\bea
 & & {\cal D}_\mu f^\mu +\partial_\lambda(M) \nabla^\lambda f =0 \\
 & & {\cal D}_\mu f +4 \Pi^\alpha f_{\alpha\mu} + \partial_\lambda(M) 
     \nabla^\lambda f_\mu =0 \\
 & & -{1\over 2} \epsilon^{\mu\nu\alpha\beta} {\cal D}_\alpha f_{5\beta}
+ \Pi^{[\mu} f^{\nu]} -\partial_\lambda(M)  \nabla^\lambda f^{\mu\nu} =0 \\
 & & \epsilon^{\mu\nu\alpha\beta} {\cal D}_\nu f_{\alpha\beta} 
 -2 \Pi^\mu f_5 + \partial _\lambda(M) \nabla^\lambda f_5^{\mu} =0 \\
 & & 2 \Pi_\mu f_5^{\mu} +\partial_\lambda(M) \nabla^\lambda f_5 =0 \\
 & & \Pi_{\mu} f^{\mu} - M f =0 \\
 & & \Pi^\mu f - M f^\mu +{\cal D}_\alpha f^{\mu\alpha} =0 \\
 & & {\cal D}_{[\mu} f_{\nu]} -4 M f_{\mu\nu} +2 \epsilon_{\mu\nu\alpha\beta}
      \Pi^\alpha f^{5\beta} =0 \\
 & & {\cal D}_\mu f_5 -2 M f_{5\mu} +2 \epsilon_{\mu\nu\alpha\beta} 
     \Pi^\nu f^{\alpha\beta} =0 \\
 & & {\cal D}_\mu f_5^{\mu} +2 M f_5 =0 
\eea
\vskip -0.1cm \noindent
where $a^{[\mu} b^{\nu]} = a^\mu b^\nu -a^\nu b^\mu$,
$\epsilon^{\mu\nu\alpha\beta}$ is the completely antisymmetrized Levi-Civita
tensor and we have defined as before $\Pi^\mu=p^\mu+\Sigma_v^\mu$, 
$M=m-\Sigma_s$, $\nabla^{\lambda} = 
d/dp_\lambda$, ${\cal D}_\mu= \partial_\mu -\partial_\lambda{\Pi_\mu}
\nabla^\lambda$.

In order to calculate the physical observables (baryon number current
$J^\mu$ and energy momentum $T^{\mu\nu}$), only the components $f=Tr(F)$
and $f^{\mu}=Tr(\gamma^\mu F)$ are necessary. The $f^\mu$ component allows 
to calculate the nucleon and vector field contributions to $J^\mu$ and 
$T^{\mu\nu}$ while $f$ gives the gap equation and the scalar field 
contributions \cite{HaMo93}.
We can obtain a kinetic-like equation by multiplying (9) with $\Pi^\mu$:
\vskip -0.4cm
\bea
\Pi_\mu\partial^\mu f -\partial_\mu(M)f^\mu-
\left[ \Pi_\mu\partial_\lambda(\Pi^\mu)-M\partial_\lambda(M) \right] 
\nabla^\lambda f =0 .
\eea
\vskip -0.1cm \noindent
A kinetic equation for $f^\mu$ can be derived from (9) and (15) as follows.
We multiply (9) with $M$ and (15) by $\Pi^\mu$ and sum the resulting 
two equations in order to eliminate the $f^{\mu\nu}$ contribution. 
We obtain in this way
\vskip -0.4cm
\bea
\!\!\!\!\Pi_\alpha\partial^\alpha\! f^\beta\!\! -\!\partial^\beta(M) f\! -\!
\left[ \Pi_\alpha\partial_\lambda(\Pi^\alpha)\!-\!M\partial_\lambda(M) \right] 
\nabla^\lambda\! f^\beta\!+\!\left[ \partial^\beta\Pi^\alpha
\!-\!\partial^\alpha\Pi^\beta \right]\! f_\alpha\! =\!0. 
\eea
\vskip -0.1cm \noindent
Equations (18) and (19) can be rewritten as
\vskip -0.4cm
\bea
\left[ \Pi^\alpha {\cal D}_\alpha -M \partial_\alpha( \Sigma_s ) \nabla^\alpha
\right] f + \partial_\mu(\Sigma_s) f^\mu =0 \\
\left[ \Pi^\alpha {\cal D}_\alpha -M \partial_\alpha( \Sigma_s ) \nabla^\alpha
\right] f^\mu +{\cal F}^{\mu\lambda} f_\lambda =0 
\eea
\vskip -0.1cm \noindent
It is straightforward to see that these equations are exactly what one obtains 
from taking the trace of (6) with $I$ and $\gamma^\mu$ respectively.
As for the mass shell equations, we obtain the one for $f$ by multiplying (14) 
by $\Pi^\mu$ and using (13)
\vskip -0.4cm
\bea
(\Pi^2 -M^2) f +\Pi_\mu {\cal D}_\nu f^{\mu\nu}=0
\eea
\vskip -0.1cm \noindent
and the one for $f^\mu$ by multiplying (10) with
$\Pi^\mu$ making use of (13) and (14) 
\vskip -0.4cm
\bea
\!\!\!\!(\Pi^2\! -\! M^2) f^\mu +\! (1/2) \epsilon^{\mu\alpha\beta\lambda} \Pi_\alpha
{\cal D}_\beta f_{5 \lambda} + M {\cal D}_\alpha f^{\mu\alpha} 
- \partial_\lambda(\Sigma_s) \Pi_\nu \nabla^\lambda f^{\mu\nu}\! =\! 0. 
\eea
\vskip -0.1cm \noindent
We notice that (22,23) differs from the traces of (7) with $I$ and 
$\gamma^\mu$. 
\par\noindent
We now proceed to show how can one recover the equations of Elze.
Let us apply derivatives\footnote{Only at this stage an operation 
is made that cannot be inverted}  to (9), obtaining $\Pi_\mu {\cal D}_\nu 
f^{\mu\nu} = \partial_\mu(\Sigma_{v\nu}) f^{\mu\nu} + {\cal O}^2$ and
$\partial_\lambda(\Sigma_s) \Pi_\nu \nabla^\lambda f^{\mu\nu}$ 
$= -\partial_\lambda(\Sigma_s) f^{\mu\lambda} +  {\cal O}^2$;
and to (15), obtaining $\epsilon^{\mu\alpha\beta\lambda} \Pi_\alpha
{\cal D}_\beta f_{5 \lambda}+ 2 M {\cal D}_\alpha f^{\mu\alpha}
= $ $\epsilon^{\mu\alpha\beta\lambda} {\cal D}_\alpha (\Pi_\beta) f_{5 \lambda}
+{\cal O}^2$. Substituting in (22,23) we see that the traces 
with $I$ and $\gamma^\mu$ of the mass shell equation (7) of Elze {\it et al.} 
are indeed recovered. The same procedure could be applied to obtain kinetic 
and mass shell equations for all components of the Wigner function. 
\par\noindent
Note that (20-23), can also be obtained directly by multiplying (3,4) 
right and left by $(\gamma.\Pi +M)$\footnote{without derivative, one can go 
back any time to (3,4) multiplying again with $(\gamma.\Pi-M)$} and take the 
sum and difference
\vskip -0.4cm
\bea
& & \! 2(\Pi^2\! -\! M^2) F \nonumber \\
& & \qquad + {i \over 2} \Pi_\mu {\cal D}_\nu\! \left\{ \sigma^{\mu\nu}\! ,F
\right\} +{i \over 2} M {\cal D}_\mu \left[ \gamma^\mu\! , F \right] 
-{i \over 2} \Pi_\mu \partial_\alpha ( \Sigma_s )\nabla^\alpha\!  
\left[ \gamma^\mu\! , F \right] =0 \\
& &  \! \left[ \Pi^\mu {\cal D}_\mu -M \partial_\alpha (\Sigma_s) \right] F 
\nonumber \\
& & \qquad + {1 \over 2} \Pi_\mu {\cal D}_\nu\! \left[ \sigma^{\mu\nu}\! , F 
\right] +{1 \over 2} M {\cal D}_\mu \left\{ \gamma^\mu \! , F \right\} 
-{1 \over 2} \Pi_\mu \partial_\alpha (\Sigma_s) \nabla^\alpha \!
\left\{ \gamma^\mu\! , F \right\}\! =\! 0 .
\eea
\vskip -0.1cm \noindent
Then, (22) and (23) are the traces of (24) with $I$ and $\gamma^\mu$.
\par\noindent
Until (24,25), we can be fully sure that no information was lost. 
It is tempting to use (6,7) since these equations take a more compact form
and have the nice feature of containing explicitly the precession
terms ${\cal F}_{\mu\nu} \left[ \sigma^{\mu\nu}, F \right]$. They will
probably give no trouble for all practical purposes. One should however 
keep in mind that the system (24,25) is more general than (6,7) as some loss
of generality is taking place in the derivation process. In any case, 
we find that the mass term 
$M \partial_\alpha(\Sigma_s) \nabla^\alpha$ is present in (20,21,25)
contrary to the claims of the authors of \cite{Smol}. We will 
examine their equations in section (\ref{sec:zub}) and show that the mass 
term discrepancy stems from a faulty mass shell condition.

\par
We now discuss a much more severe restriction commonly imposed on the mass
shell, which consists of approximating (7) by $(\Pi^2-M^2) F=0$.
If $f$ and $f^{\mu}$ are remaining on the mass shell $\Pi^2-M^2=0$,
equations (20,21) would represent a self contained system for the evolution 
of $f$ and $f^{\mu}$. From (22,23) we see that in general however they
deviate from this mass shell by terms involving the derivatives
of $f^{\mu\nu}$ and $f_5^\mu$. Only in symmetric nuclear matter in a state 
of (homogeneous) equilibrium, and when the system is unpolarized, does the 
stationary solution reduce $F=(\gamma.\Pi+M)\delta(\Pi^2-M^2)\varphi$. 
Unfortunately, very often it is assumed that the spinor structure remains 
unchanged out of equilibrium, ($F=f+f_\mu \gamma^\mu$, 
$0=f^{\mu\nu}=f_5=f_{5\mu}$). Then the system (8-17) simplifies, namely: 
Eqs. (11,12,16,17) contain no information.
Due to the assumption $f^{\mu\nu}=0$, Eq. (14) gives $f^{\mu}=(\Pi^{\mu}/M) f$
while Eq. (13) returns the mass shell condition $(\Pi^2-M^2)f=0$. Eq. (10)
gives $\Pi^{[\mu} f^{\nu]}=0$ which is trivial with $f^{\mu}=(\Pi^{\mu}/M) f$.
We are left with Eqs. (8,9,14), and also (18,19) which were derived from the 
former. Then, either
\par\hskip 0.2cm * by replacing $f^{\mu}$ by $(\Pi^{\mu}/M) f$ in Eq. (8),
\par\hskip 0.2cm * or multiplying (9) by $\Pi^\mu$ and replacing $f^{\mu}$,
\par\hskip 0.2cm * or replacing $f^{\mu}$ in Eq. (19) which was obtained from
(9) and (15),
\par\noindent
we arrive at
\vskip -0.4cm
\bea
\Big\{ \Pi^\mu\partial_\mu -\big[ \Pi^\mu\partial_{\lambda}(\Pi_\mu)
-M\partial_\lambda(M) \nabla^\lambda\big]\Big\}({f \over M}) =0.
\eea
\vskip -0.1cm
This equation might be further transformed by taking the change of
variables $(x,p)\rightarrow(x,\Pi)$ so that we obtain
\vskip -0.4cm
\bea
\Big\{ \Pi^\mu\partial_\mu - \Pi^\mu{\cal F}^{\mu\lambda}{\partial \over
\partial \Pi^{\lambda}} -M\partial_\lambda(M) {\partial \over
\partial \Pi^{\lambda}}\big]\Big\}({f \over M}) =0.
\label{27}
\eea
\vskip -0.1cm \noindent
with ${\cal F}^{\mu\lambda}=\partial^\mu(\Sigma_v^\lambda)
-\partial^\lambda(\Sigma_v^\mu)$.
This is the equation which is extensively used in numerical simulations of 
heavy ion collisions with a Boltzmann-Uehling-Uhlenbeck collision term on 
the right hand side.
\par\noindent
Since Eq. (\ref{27}) was arrived at by three different methods, we could think
that the result is indeed safe. However we still not have used {\it all}
the information available in the system (8-17). For example, we could 
have replaced $f^{\mu}$ by $\Pi^\mu/M f$ and take $f^{\mu\nu}=0$ 
in Eq. (15) directly and obtain
\vskip -0.4cm
\bea
\Pi^{[\mu}\partial^{\nu]}({f\over M})-\Pi^{[\mu}\partial_\lambda(\Pi^{\nu]})
\nabla^\lambda ({f\over M})=0. 
\eea
\vskip -0.1cm \noindent
Strictly speaking, we simply do not have a justification to put 
$f^{\mu\nu}=0$. This term has no reason to vanish out of equilibrium.
It could for example acquire contributions like ${\cal F}^{\mu\nu} f$,
$\Pi^{[\mu}\partial^{\nu]}(M) f$, or time-reversal invariance breaking terms 
like $U^{[\mu} f^{\nu]}$. 

\vskip -0.1cm \noindent
\subsection{Vlasov equation from the ``Zubarev'' method}
\label{sec:zub}
\vskip -0.1cm \noindent

In this alternative derivation \cite{Smol} (``Zubarev formalism''), after 
defining a proper time $\tau=x^\mu u_\mu$, one puts the Dirac equation into 
Hamiltonian form
\vskip -0.4cm
\bea
 i u^\mu\partial_\mu \psi = i {\partial \psi\over\partial \tau}=H \psi \quad
{\rm with} \
 H &=& -i \sigma^{\mu\nu} u_{\mu}\partial_\nu+m \gamma.u+\gamma.u \Sigma
e^{-i \Delta \over 2}  \nonumber 
\eea
\vskip -0.1cm \noindent
The conjugate equation for $\overline \psi$ reads
\vskip -0.4cm
\bea
i u^\mu\partial_\mu \overline\psi =i {\partial \overline\psi \over
\partial\tau} =\overline\psi \widetilde H \quad {\rm with} \ 
\widetilde H &=& i \sigma^{\mu\nu}u_\mu\overleftarrow{\partial_\nu}
-m \gamma.u-e^{i \Delta \over 2} \widetilde \Sigma \gamma.u.
\eea
\vskip -0.1cm \noindent
We have $\widetilde H= - \gamma_0 H^\dagger \gamma_0$ and
$\widetilde \Sigma= \gamma_0 \Sigma^\dagger \gamma_0 = \Sigma$. 
Acting on the definition of $F$ with $u.\partial F=...$ and making 
these replacements, we arrive at\footnote{Note that this can be put
in the form of the Von Neumann equation $i \partial_0 F = [H,F]$ 
if we choose the $0$ coordinate along $u^\mu$, so that $\widetilde
\Sigma \gamma.u = \gamma.u \Sigma$ and $\widetilde H = -H$ in the 
mean field approximation.}    
\vskip -0.4cm
\bea
u.\partial F=\int d^4R e^{-i p.R} \left[ -i \overline \psi \widetilde H 
\otimes \psi -i \overline \psi \otimes H \psi \right] 
= -i (H F +F \widetilde H).
\eea
\vskip -0.1cm \noindent
Choosing now $u^\mu=\Pi^\mu /\sqrt{\Pi^2}$, one arrives at the ``kinetic
equation'' of Smolyansky {\it et al.} \cite{Smol}. 
\vskip -0.4cm
\bea
& \Pi.\partial F +{1 \over 2} \Pi_\mu \partial_\nu [\sigma^{\mu\nu},F]
+i M [\gamma.\Pi,F] - \Pi^\mu \partial_\alpha (\Sigma_{v \mu}) \nabla^\alpha F
& \nonumber \\
&-{1 \over 2} \Pi_\mu \partial_\alpha (\Sigma_{v \nu}) \left[ \sigma^{\mu\nu}, 
\nabla^\alpha F \right] -{1 \over 2} \partial_\alpha(\Sigma_s)
\left\{ \gamma.\Pi, \nabla^\alpha F \right\} =0
\label{eq:smolkin}
\eea
\vskip -0.1cm \noindent
One can easily see that the same equation can be obtained by multiplying
the eqs. (3,4) by $\gamma.\Pi$ from the left and from the right
and taking the sum.

However, we still have an other equation, obtained by multiplying on 
the right and left hand side and subtracting, which is the corresponding
constraint equation. It reads 
\vskip -0.4cm
\bea
& {i \over 2} \Pi_\mu \partial_\nu \left\{ \sigma^{\mu\nu}, F \right\}
+2 \Pi^2 F -M \left\{ \gamma.\Pi, F \right\}  & \nonumber \\
& -{i \over 2} \Pi_\mu
\partial_\alpha(\Sigma_{v \nu}) \left\{ \sigma^{\mu\nu}, \nabla^\alpha F 
\right\} -{i \over 2} \left[ \gamma.\Pi , \partial_\alpha(\Sigma_s) 
\nabla^\alpha(F) \right]=0 &
\label{eq:smolmas}
\eea
\vskip -0.1cm \noindent
Until now, the system (\ref{eq:smolkin},\ref{eq:smolmas}) is fully equivalent 
to the original set of equations (3,4)\footnote{Since we can always recover 
the initial set of equations by multiplying a second time with $\gamma.\Pi$ 
and use $(\gamma.\Pi)^2=\Pi^2$}. However, the second 
``constraint'' equation is not taken into account by Smolyansky {\it et al.}
 who replace it by the mass shell condition $\Pi^2=M^2$, 
although we can see from the explicit expression that
the $M^2$ term is missing and that it contains instead dynamics at the same 
order as the other equation. 
Moreover, one can see that in the free case, the well known kinetic
equation $p.\partial F=0$, is not recovered from
(\ref{eq:smolmas})
alone \footnote{It can be 
shown that the equations $p.\partial F={\cal O}^2$ and $(p^2-m^2)F ={\cal O}^2$
can be recovered from the system (31),(32) by submitting it to a series of 
manipulations which amounts to return back to the system (3,4) by multiplying 
(31) and (32) by $\gamma.\Pi$, and taking the sum and difference, and then 
apply the technique exposed in the previous section}.
This example illustrates that some care has to be paid that the seemingly 
obvious mass shell condition should be obtained from the calculation, 
less some part of the dynamics may be neglected by imposing this condition 
beforehand.

\vskip -0.8cm
\subsection{Vlasov equation: discussion}
\vskip -0.2cm

There are several ways one can derive an equation of the form 
$\Pi.\partial F + D_K[F] =0$, each method of derivation leading to a different 
set of additional terms $D_K[F]$. To this equation should be associated
the corresponding mass shell condition which also contains dynamical effects
$(\Pi^2-M^2)F+ D_M[F] =0$. If the usual mass shell condition were 
imposed in the form $(\Pi^2-M^2)F=0$ from outside, terms $D_M[F]$ of the
same order as $D_K[F]$ would be neglected, leading to a wrong description 
of the dynamics. The Zubarev method is particularly treacherous in this 
respect, since the neglected terms are of order zero in the gradient 
expansion. The method developed by Elze {\it et al.} on the other hand
permits to minimize the importance of the $D_M[F]$ terms so that these are of 
second order in the gradient expansion if the interactions are switched off.
When the mean field is present, however, $D_M[F]$ remains of first
order and has to be considered. 

\vskip -0.1cm \noindent
\section{Pseudoscalar {\it vs.} Pseudovector mean fields}
\vskip -0.1cm \noindent

In order to illustrate the highly nontrivial structure of the mass shell 
constraint in the general case, we will compare here the stationary,
homogeneous solution in the presence of a nonvanishing pseudoscalar or 
pseudovector mean field. The equations (3,4) now read 
$(\gamma.p -m - \Sigma) F(x,p) = 0 = F(x,p) (\gamma.p-m -\Sigma)$
The solution of these equations is found to be  \cite{MoMo00}
\vskip -0.4cm
\bea
F(x,p) = P_L(x,p)\ Q(x,p)\ P_R(x,p)\ \delta(D)
\eea
\vskip -0.1cm \noindent
Here, $P_L(x,p)$ and  $P_R(x,p)$ are matrices in the Clifford 
algebra such that $(\gamma.p -m - \Sigma)\ P_R(x,p) = D =
P_L(x,p)\ (\gamma.p -m - \Sigma)$, $D=det(\gamma.p-m -\Sigma)$ 
being a c-number. The Dirac $\delta$ function provides for the 
cancelation of the products $(\gamma.p -m - \Sigma) F$ and 
$ F (\gamma.p-m -\Sigma)$. $Q(x,p)$ is an arbitrary  matrix in 
the Clifford algebra and contains the thermodynamic averages. 
It can be shown that $P_R=P_L=P$, and $P$ can be interpreted as 
a projector on the positive-energy state on the mass shell $D=0$. 

\vskip -0.1cm \noindent
\subsection{Scalar, vector and pseudoscalar mean fields}
\vskip -0.1cm \noindent

We assume  here that the mean field self energy is of the form: 
$\Sigma=(\Sigma_s +\Sigma_{v\mu} \gamma^\mu +\Sigma_5 \gamma_5)$.
The stationary homogeneous solution to the Vlasov system of equations 
is now found to be\footnote{The reader may easily convince himself that
inserting the expression for $f^{\mu\nu}$ in that of $f_5^\mu$ leads 
to an identity}
\vskip -0.4cm
\bea
& (p_*^2-m_*^2 -\Sigma_5^2)f =0 \quad ; \qquad 
f^{\mu}=\displaystyle{p_*^{\mu} \over m_*} f \quad ; \qquad 
f_5=\displaystyle{\Sigma_5 \over m_*} f \quad ; & \nonumber \\
& f_5^{\mu} = \displaystyle{1 \over m_*} \epsilon^{\mu\nu\rho\lambda} 
p^*_{\nu} f_{\rho\lambda} \quad ; \qquad f^{\mu\nu} = 
\displaystyle{ \Sigma_5 p_*^{[\mu} f_5^{\nu]} 
+ m_* \epsilon^{\mu\nu\rho\lambda} p^*_{\rho} f_{5\lambda} \over 
\Sigma_5^2 +m_*^2} & 
\eea
\vskip -0.1cm
\noindent with again $p_*^{\mu}=p^{\mu}+\Sigma^{v\mu},\ m_*=m-\Sigma_s$.
The solution factorizes as 
\vskip -0.4cm
\bea
F(x,p)=P_{\oplus} \left[ \Lambda_{\oplus} \delta(p_*^2-m_*^2-\Sigma_5^2)
\varphi(x,p) \right] P_{\oplus}
\eea
\vskip -0.1cm
\noindent with generalized definitions of two projection operators:
The projector on the positive energy state of the mass
shell $p_*^2-m_*^2-\Sigma_5^2=0$
\vskip -0.4cm
\bea
 P_{\oplus} = 1+ \displaystyle{p_*^{\mu} \over m_*} \gamma_{\mu}
+i \displaystyle{\Sigma_5 \over m_*} \gamma_5
\eea
\vskip -0.1cm \noindent
and the projector on the local spin direction
\vskip -0.4cm
\bea
\Lambda_{\oplus}= ( 1+\gamma_5 \gamma_{\mu} {\cal S}^{\mu})/2; \ \ 
{\cal S}^{\mu} =\displaystyle{m_* f_5^{\mu} \over (m_*^2+\Sigma_5^2) \varphi}
\eea
\vskip -0.1cm
\noindent with $f=\delta(p_*^2-m_*^2-\Sigma_5^2) \varphi$ and the constraint 
$p^*_\mu {\cal S}^{\mu}=0$
\par\noindent In this case, we have four independent distribution functions:
$f$ and the three components of $f_5^\mu$ orthogonal to $p_*^\mu$.

\vskip -0.1cm \noindent
\subsection{Scalar, vector and pseudovector mean fields}
\vskip -0.1cm \noindent

Let's assume that $\Sigma=(\Sigma_s +\Sigma_{v\mu} \gamma^\mu 
+\Sigma_{5\mu} \gamma_5 \gamma^{\mu})$. We find that  the stationary 
homogeneous solution of the Vlasov
equation contains only one independent distribution function (the scalar one
$f$). The other components are given by
\vskip -0.4cm
\bea
& f^{\mu}\! =\! {\cal A} \displaystyle{\Sigma_{5\mu} \over m_*}\! +\! {\cal B} 
\displaystyle{p_*^{\mu} \over m_*} \ ; \ 
 f_5^{\mu}\!=\! {\cal C} \displaystyle{\Sigma_{5\mu} \over m_*} \!+\! {\cal D} 
\displaystyle{p_*^{\mu} \over m_*} \ ; \ f_5\!=\!0  \ ; \ 
 f^{\mu\nu}\!= \!\epsilon^{\mu\nu\rho\lambda} \displaystyle{p^*_{\rho} 
\Sigma_{5\lambda} \over m_*^2} {\cal E} & 
\eea
\vskip -0.1cm \noindent
The expressions of ${\cal A,B,C,D,E}$ are given in the appendix. 
The mass shell condition is given by $-m_* f +p^*_{\alpha} f^{\alpha} 
- \Sigma_{5\alpha} f_5^{\alpha} =0$ or equivalently $D f=0$ with
\vskip -0.4cm
\bea
D & = & \left[ (1-p_*^2/m_*^2-\Sigma_{5\alpha}\Sigma_5^\alpha/m_*^2)^2
-4 (p^*_{\alpha}\Sigma_5^{\alpha}/m_*^2)^2 
+4 \Sigma_{5\alpha}\Sigma_5^\alpha/m_*^2\right]  \\
& =& m_*^4\! -\!2 m_*^2(p^*_{\alpha}\!+\!\Sigma_{5\alpha})
(p^*_{\alpha}\!-\!\Sigma_{5\alpha})\! +\! (p^*_{\alpha}\!-\!\Sigma_{5\alpha})
(p^*_{\beta}\!-\!\Sigma_{5\beta})(p^*_{\alpha}\!+\!\Sigma_{5\alpha})
(p^*_{\beta}\!+\!\Sigma_{5\beta}) \nonumber
\eea
\vskip -0.1cm \noindent
\label{EQmassshellPV}
The solution can again be expressed in a compact form with the help
of a generalized projector $P_{\oplus}$
\vskip -0.4cm
\bea
F(x,p) & =&  P_{\oplus} \delta(D) \varphi(x,p) 
\eea
\vskip -0.8cm
\bea
P_{\oplus} & = &  \!\displaystyle{
(m_*\!-\!p^*_{\mu}\gamma^{\mu}\!-\!\Sigma_{5\mu}\gamma_5\gamma^{\mu})
(m_*\!-\!p^*_{\nu}\gamma^{\nu}\!-\!\Sigma_{5\nu}\gamma_5\gamma^{\nu})
(m_*\!-\!p^*_{\rho}\gamma^{\rho}\!-\!\Sigma_{5\rho}\gamma_5\gamma^{\rho}) \over
2 m_* (m_*^2\!-\!p_*^2 +\Sigma_{5\alpha}\Sigma_5^{\alpha})} \nonumber
\eea
\vskip -0.1cm \noindent
on the mass shell $D=0$ defined as above (eq. 39).
It is obvious that the physics described by the solution in a pseudoscalar
field will be very different from the physics in a pseudovector field.
In the case of the Walecka model for example, it will thus not be a 
simple matter of convenience to choose PS or PV coupling of the pion field.

\vskip -0.3cm \noindent
\section{Conclusion}
\vskip -0.1cm \noindent

On the basis of two examples, we have wished to stress the importance of
taking into account the mass shell condition properly. Only in very simple
cases does this condition take the simple form $(\Pi^2-M^2)F=0$, in general
however it is highly non trivial.
\par\noindent
In the case of the $\sigma$ - $\omega$ model of nuclear matter, it was shown
that one cannot separate clearly the ``kinetic'' and ``mass shell'' 
conditions in general. The mass shell condition always contains some dynamical 
contribution in the interacting case. Overlooking this fact may even lead 
to wrong dynamics.
\par\noindent
In the case of a system with pseudoscalar or pseudovector coupling,
we have solved the stationary case and shown that the mass shell condition 
acquires a rich spinor structure. The PS or PV coupling lead to very different 
solutions.

\vskip -0.3cm \noindent
\section*{Appendix}
\vskip -0.1cm \noindent

The full expression of the coefficients entering the PV result is:
\vskip -0.4cm
\bea
{\cal A} & = & \left[ (p^*_{\alpha} \Sigma_5^{\alpha}/m_*^2\ 
(f+p^*_\alpha f^{\alpha}/m_*) -(p_*^2/m_*^2-1) \Sigma_{5\alpha} 
p_*^{\alpha}/m_*^2 f \right] / {\cal F} \nonumber \\
{\cal B} & = & \left[ (p_*^2/m_*^2-1)(\Sigma_{5\alpha} f_5^{\alpha}/m_*-f)
-(\Sigma_{5\alpha} p_*^{\alpha}/m_*^2)^2 f \right]/ {\cal F} \nonumber \\
{\cal C} & = & \left[ (1+\Sigma_{5\alpha}\Sigma_5^\alpha/m_*^2)\ (f+p^*_\alpha 
f^{\alpha}/m_*) -(\Sigma_{5\alpha}p_*^{\alpha}/m_*^2)^2 f \right]/ {\cal F} 
\nonumber \\
{\cal D} &= &\left[ (\Sigma_{5\alpha}p_*^{\alpha}/m_*^2)\Sigma_{5\alpha} 
f_5^{\alpha}/m_*-2 f -\Sigma_{5\alpha}\Sigma_5^\alpha/m_*^2 f) \right]/  
{\cal F} \nonumber \\
{\cal E}& = & -({\cal A} + {\cal D})/2 \nonumber \\
 & = & \left[ (1+\Sigma_{5\alpha}\Sigma_5^\alpha/m_*^2)
(1+(\Sigma_{5\alpha}p_*^{\alpha}/m_*^2)^2-p_*^2/m_*^2 \right]
/( {\cal F} {\cal G)} \nonumber \\
{\cal F} & = &  (\Sigma_{5\alpha}p_*^{\alpha}/m_*^2)^2 +(1-p_*^2/m_*^2)
(1+\Sigma_{5\alpha}\Sigma_5^\alpha/m_*^2) \nonumber \\
\Sigma_{5\alpha} f^{\alpha} & =& 
\left[ 2 (p^*_{\alpha}\Sigma_5^{\alpha}/m_*^2)^2
\!\!-\! \Sigma_{5\alpha}\Sigma_5^{\alpha}/m_*^2 (1\!+\!p_*^2/m_*^2 \!+\!
\Sigma_{5\alpha}\Sigma_5^{\alpha}/m_*^2) \right]\! /{\cal G} m_* f \nonumber \\
p^*_{\alpha}f^{\alpha} & =&\! \left[ -2 (p^*_{\alpha}\Sigma_5^{\alpha}/m_*^2)^2
\!+\!p_*^2/m_*^2\ (p_*^2/m_*^2\!-\! 1\! +\!\Sigma_{5\alpha}
\Sigma_5^\alpha/m_*^2)\right]\! /{\cal G} m_* f \nonumber \\
{\cal G} & = & p_*^2/m_*^2-1-\Sigma_{5\alpha}\Sigma_5^{\alpha}/m_*^2  
\nonumber
\eea


\vskip -0.1cm
\section*{References}
\begin{thebibliography}{99}

\bibitem{FloHueKl96} W. Florkowski, J. H{\"u}fner, S.~P. Klevansky, 
and L. Neise, \Journal{\em Ann. Phys. (N.Y.)}{245}{445}{1996}

\bibitem{MrHe94}
S. Mrowczynski and U. Heinz, \Journal{\em Ann. Phys. (N.Y.)}{\bf 1}{1}{1994}

\bibitem{HaMo92} R. Hakim, L. Mornas, P. Peter, and H. Sivak, 
\Journal{\PRD}{46}{4603}{1992}


\bibitem{Flo98} W. Florkowski, \Journal{\em Eur.Phys.J. }{A2}{77}{1998}.

\bibitem{helium} A. Meyerovich, \Journal{Physica B}{169}{183}{1991} \\
I. Jeon and W. Mullin, \Journal{J. Phys. (Paris)}{49}{1691}{1988} \\
C. Lhuilier and F. Lalo{\"e}, \Journal{J. Phys. (Paris)}{43}{197,225}{1982}

\bibitem{MoMo00} L. Mornas, K. Morawetz, {\em to be published} 

\bibitem{EG87} H.T. Elze, M. Gyulassy, D. Vasak, U. Heinz, H. Stocker and
   W. Greiner,  \Journal{\em Mod. Phys. Lett.}{A2}{451}{1987}

\bibitem{HaMo93} R. Hakim and L.Mornas, 
\Journal{\em Phys. Rev.}{C47}{2846}{1993} 

\bibitem{SW86} B.D. Serot, J.D. Walecka, ``The Relativistic Nuclear Many-Body 
Problem'', Adv. in Nucl. Phys., {\bf 16}, J.W. Negele and E. Vogt Edts.
( Plenum, New York, 1986).

\bibitem{Smol} V.~D. Toneev, A.~V. Prozorkevich, and S.~A. Smolyansky, 
\Journal{\em Heavy Ion Physics}{3}{37}{1996} \\
S.G. Mashnik, A.V. Prozorkevich, S.A. Smolyansky, G. Maino
\Journal{\em Nuovo Cim.} {109A}{1699}{1996} 

\end{thebibliography}
\end{document}